\documentclass[10pt,article]{dis03}
\usepackage{epsf,amsmath,graphicx}

\newcommand{\etal}{{\it et al.}}
\newcommand{\hpp}{H^{\pm \pm}}
\newcommand{\pbi}{\,\text{pb}^{-1}}
\newcommand{\txtd}{\text{d}}

\begin{document}

\title{Multi-lepton events at HERA}

\author{
  Andrea Parenti\\
  (on behalf of H1 and ZEUS Collaborations)\\
  Dipartimento di Fisica ``G. Galilei'' and INFN\\
  via Marzolo 8, 35131 Padova, Italy\\
  E-mail: parenti@pd.infn.it
}

\maketitle

\begin{abstract}
\noindent
 Events containing high-$P_T$ multi-leptons were sought
 at HERA by the H1 and ZEUS collaborations.
 Experimental data were compared to simulations of
 $\gamma \gamma \rightarrow l^+ l^-$ processes,
 and H1 extracted the total and differential cross sections
 of the processes.

 The doubly-charged Higgs, which decays into lepton pairs,
 was analysed by H1 as a possible source of the multi-electron
 events, which have been observed at high invariant masses.
\end{abstract}

\section{Introduction}
 The high centre-of-mass energy of HERA (around 300~GeV)
 offers the possibility of
 producing events with two (or more) leptons at high transverse energy ($E_T$).
 The production rate of these events can be accurately predicted within
 the Standard Model (SM).
 The H1 and ZEUS collaborations compared the data collected by their detectors
 \cite{h1det,zeusdet} to the SM, in order to detect any deviation.

\section{Lepton-pair production at HERA}
\begin{figure}
[!thb]
  \begin{center}
    \includegraphics[height=3.8cm]{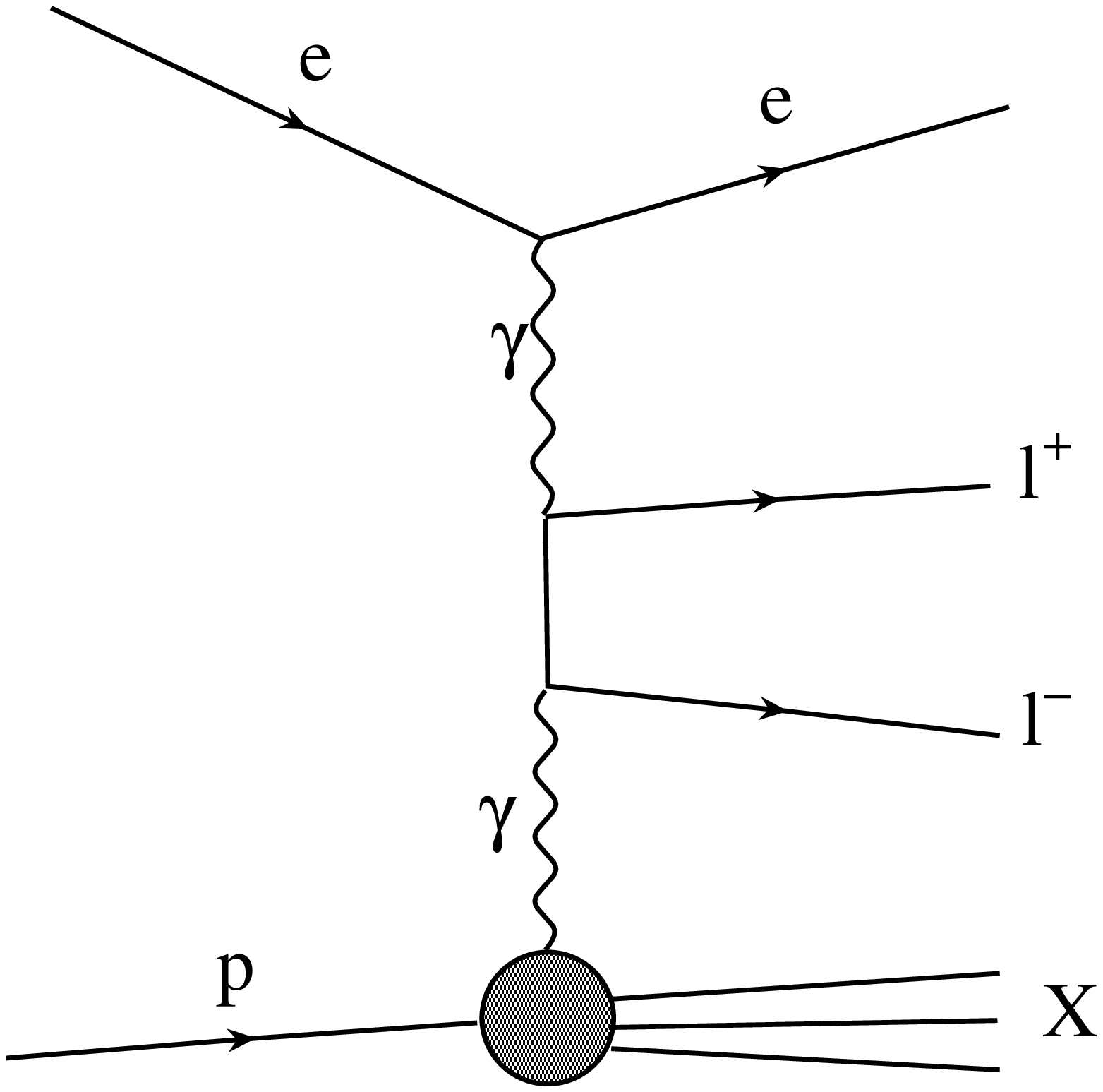}
    \includegraphics[height=3.8cm]{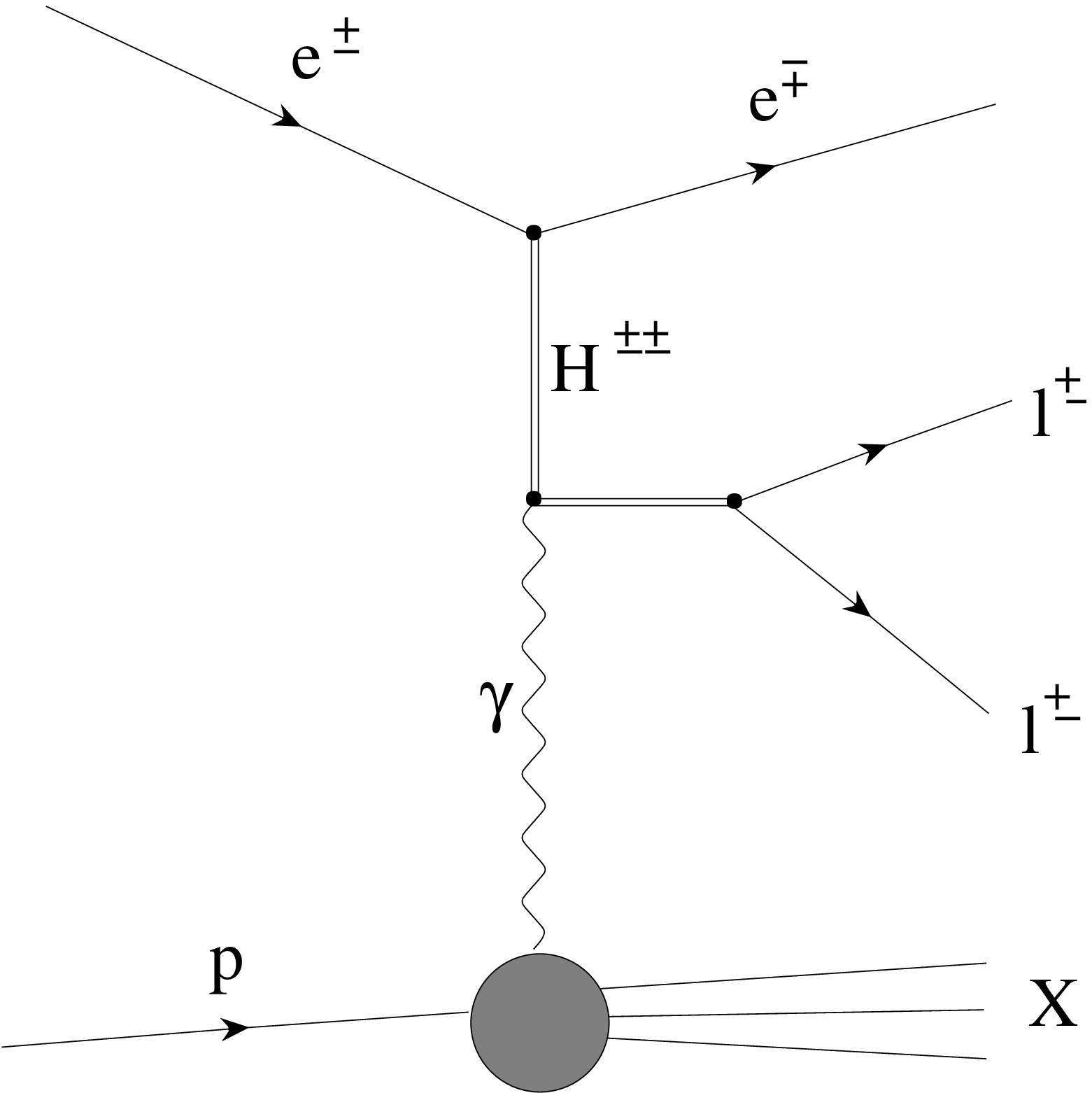}
    \caption{Lepton pair production via the $\gamma \gamma$ process (left);
      this is the major contribution to the $l^+l^-$ production.
      Production of $\hpp$ (right) and its subsequent decay into
      two like-sign leptons; only one of the possible diagrams is shown.}
    \label{fig-feynm}
  \end{center}
\end{figure}

 At HERA the dominant process which produces high-$E_T$ lepton pairs is the
 non-resonant electroweak process $\gamma \gamma \to l^+ l^-$
 (Fig.~\ref{fig-feynm}, left).
 Processes beyond the Standard Model may contribute to
 lepton pair production; the supersymmetric left-right models (SUSYLR,
 \cite{susylr}) predict the existence of a doubly-charged Higgs
 ($\hpp$), which can decay into a pair of like-sign charged leptons
 (Fig. \ref{fig-feynm}, right).

 The $\gamma \gamma$ process was simulated using the GRAPE Monte Carlo
 \cite{grape}, whereas the $\hpp$ signal was generated
 by CompHEP \cite{comphep}.

\section{Search for di-muon events}
 At H1, muon candidates were selected from tracks measured
 in the central tracker, which were linked to tracks found
 by the muon detector, or to a minimal ionising particle (MIP) deposit
 in the calorimeter (for low momentum tracks).

 The search for di-muons \cite{h1dimu} was carried out
 in the phase space defined by:
 \mbox{$M_{\mu , \mu} > 5$ GeV},
 \mbox{$P^{\mu 1}_T > 2.0$ GeV},
 \mbox{$P^{\mu 2}_T > 1.75$ GeV},
 \mbox{$20^\circ < \theta_\mu < 160^\circ$},
 \mbox{$D^\mu_\text{Track,Jet} > 1$}
 (or \mbox{$D^\mu_\text{Track,Jet}> 0.5$ if $P^\mu_T > 10$ GeV})
 \footnote{$D^\mu_\text{Track,Jet}$ is the distance of the muon
   to the nearest track or jet in the pseudorapidity-azimuth-plane.}.

 GRAPE was found to describe well all data distributions.
 The total cross section measured by H1 is
 \mbox{$\sigma=46.5 \pm 1.3 \pm 4.7$~pb} (46.2~pb expected from SM);
 the inelastic component was extracted by tagging the proton
 remnant $X$ in the forward detectors \cite{h1elamu}:
 \mbox{$\sigma_\text{inel}=20.8 \pm 0.9 \pm 3.3$~pb} (21.5 predicted by SM).

 At ZEUS, muon candidates were selected requiring a track
 in the central tracker (CTD) pointing to a MIP in the calorimeter (CAL).
 In its analysis \cite{zeus_emu},
 the ZEUS collaboration required two isolated,
 energetic \mbox{($P_T^\mu>5$ GeV)} muons in the central region
 \mbox{($20^\circ < \theta_\mu < 160^\circ$)}
 separated by an angle \mbox{$\Omega < 174.2^\circ$};
 a comparison of the data to the GRAPE Monte Carlo shows
 a reasonable agreement (see Tab.~\ref{tab-summary}).

\begin{figure}[!thb]
  \begin{center}
    \includegraphics[height=3.6cm]{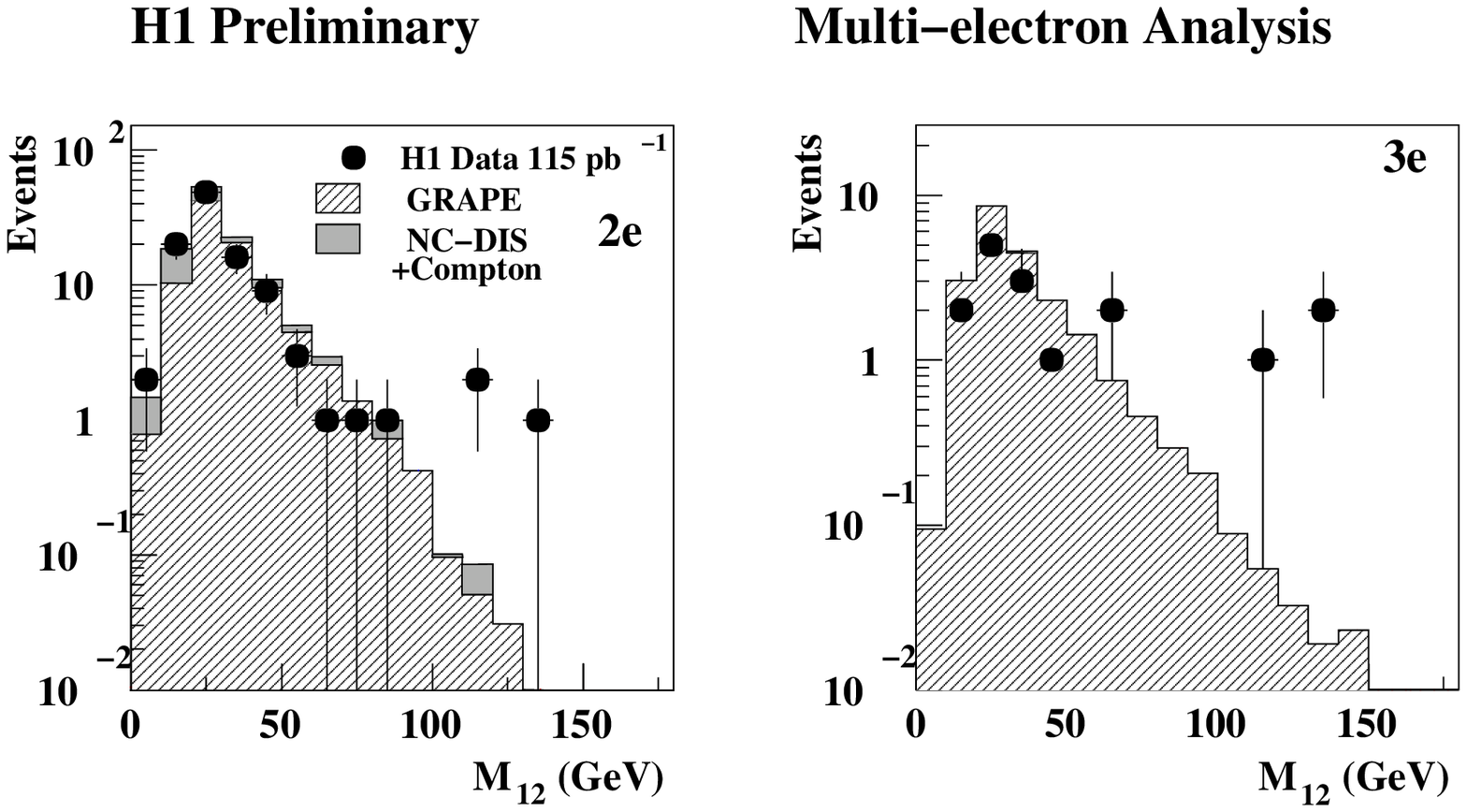}
    \includegraphics[height=3.6cm]{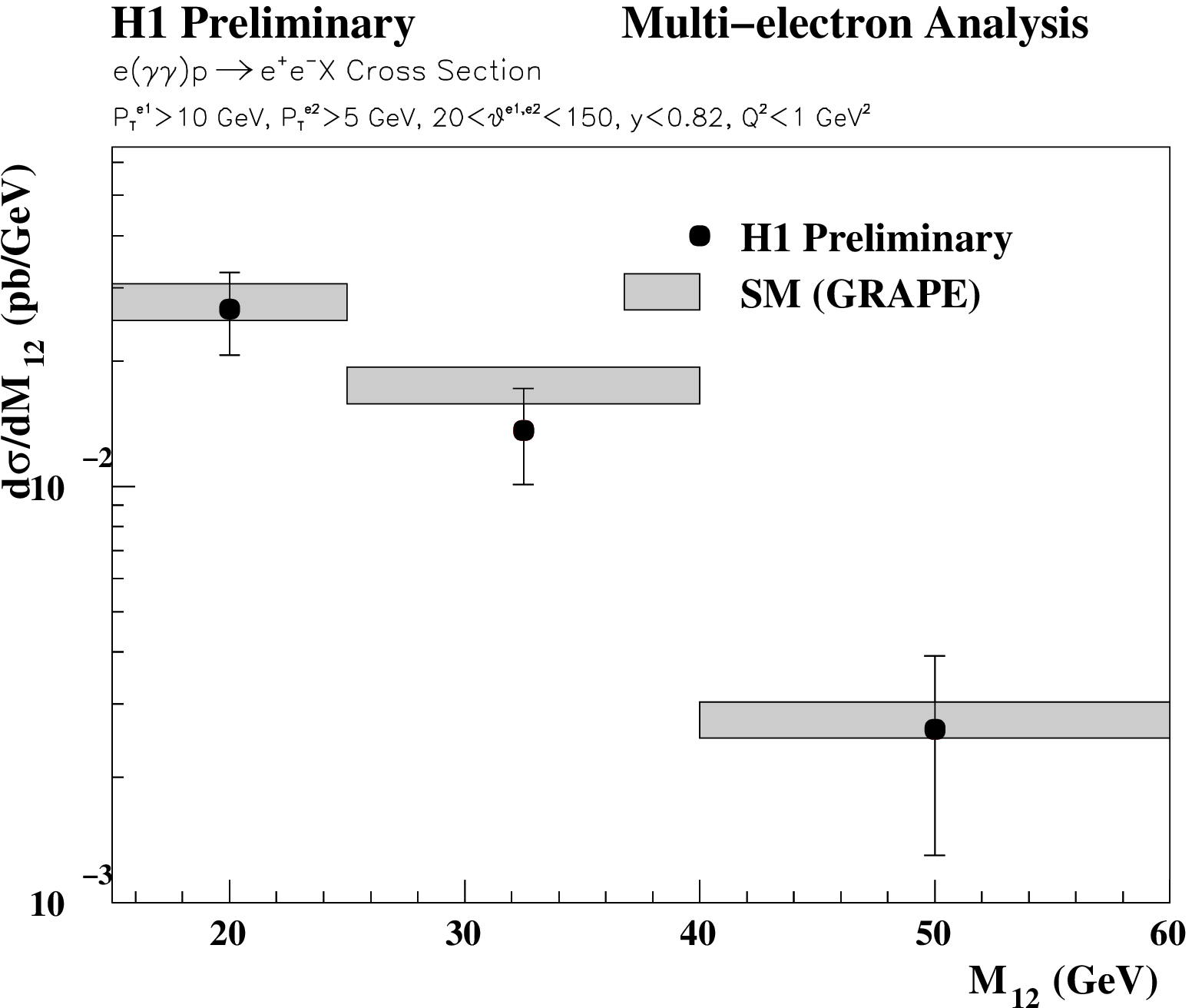}
   \caption{Invariant mass $M_{12}$ for ``2e'' (left) and
     ``3e'' (centre) samples at H1. Differential cross section
     $\txtd \sigma / \txtd M_{12}$  (right) from $\gamma \gamma$ sample.}
  \label{fig-h12e}
  \end{center}
\end{figure}
\section{Search for multi-electron events}
 H1 performed an electron identification of clusters detected
 in the electromagnetic calorimeter;
 for candidates in the central region
\footnote{In H1 analysis electrons were classified,
  depending on their polar angle, as
  forward \mbox{($5^\circ < \theta_e < 20^\circ$)},
  central \mbox{($20^\circ < \theta_e < 150^\circ$)} or
  backward \mbox{($150^\circ < \theta_e < 175^\circ$)}.
}
 a track matched in momentum and position was required.

 In the H1 analysis \cite{h12e} all electrons were asked to be
 isolated \mbox{($D^e_\text{Track}>$ 0.5)}
 and energetic: \mbox{$E_e>5$ GeV} (central and rear regions)
 or \mbox{$E_e>10$ GeV} (forward).
 The electrons were sorted in transverse energy:
 \mbox{$E_T^{e1} > E_T^{e2}$}.
 At least two central electrons were required, one with 
 \mbox{$E_T>10$ GeV} and one with \mbox{$E_T>5$ GeV}.

 The events were classified as ``2e'' (two observed electrons) or ``3e''.
 The $\gamma \gamma$ sample was defined as: ``2e'', opposite-sign
 electrons and  \mbox{$E-P_z<45$ GeV}
 (so that only events where the scattered electron is lost
 in the beam pipe are kept).
 Fig.\ref{fig-h12e}, left and centre, shows the distribution of
 invariant mass of the highest $E_T$ di-electrons, $M_{12}$.
 On the right the differential cross section $\txtd \sigma / \txtd M_{12}$
 extracted from the $\gamma \gamma$ sample is shown.
 Good agreement is found between data and SM at low $M_{12}$
 (see also Tab. \ref{tab-summary}),
 an excess is observed at high mass: 3 events in ``2e'' sample
 ($0.25 \pm 0.05$ expected) and 3 events in ``3e'' ($0.23 \pm 0.04$ expected).

 The doubly-charged Higgs was analysed by H1 \cite{h1hgs}
 as a possible explanation of this excess;
 further cuts were applied to the sample, in order
 to optimize the sensitivity to $\hpp$. Only one event survived the cuts,
 although the efficiency on the signal is quite high (20-45\%).
 Therefore the 6 events observed by H1 are unlikely to originate from
 $\hpp$ production, and limits were set on this process;
 assuming a coupling $h_{ee}=0.3$ and a branching ratio $BR(\hpp \to e^\pm
 e^\pm)=1/3$, a doubly-charged Higgs with $M_\hpp < 102$~GeV was excluded.

 At ZEUS electron candidates were selected by an algorithm which combines
 CTD information with electromagnetic clusters measured by the CAL;
 for candidates in the central region
 an energetic ($P>5$~GeV) track matched in position was required.

\begin{figure}[!thb]
  \begin{center}
    \includegraphics[height=3.5cm]{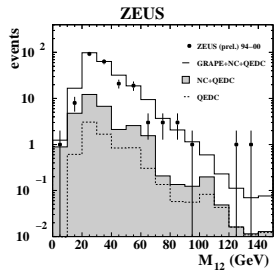}
    \caption{Invariant mass distribution of multi-electron events at ZEUS.}
    \label{fig-zeusee}
  \end{center}
\end{figure}

 In the ZEUS analysis \cite{zeus_emu},
 all electrons were asked to be isolated and energetic:
 \mbox{$E_e>10$~GeV} (forward region
\footnote{
  Electrons were classified by ZEUS, depending on their polar angle, as
  forward \mbox{($5^\circ < \theta_e < 17^\circ$)},
  central \mbox{($17^\circ < \theta_e < 164^\circ$)} or
  backward \mbox{($164^\circ < \theta_e < 175^\circ$)}.
})
 or \mbox{$E_e>$ 5~GeV} (central-rear regions).
 The events were accepted if they contain at least 2
 central electrons, one with $E_T>10$~GeV.
 The SM describes well the data (Fig. \ref{fig-zeusee}
 and Table \ref{tab-summary}). No excess was observed for $M_{12}>100$~GeV:
 2 events were found with ``2e'' topology ($0.77 \pm 0.08$ predicted by SM)
 and 0 in ``3e'' sample ($0.37 \pm 0.04$ in SM).

\section{Conclusions}
 HERA collisions were analysed by H1 and ZEUS in the search for
 multi-lepton events. Data are well described by the
 simulation of \mbox{$\gamma \gamma \to l^+ l^-$} process,
 except for the high-mass region in multi-electron events at H1,
 where an excess is observed.
 The doubly-charged Higgs, decaying into di-leptons,
 was investigated as a possible source of the excess,
 and found to be unable to explain it.

\begin{table}
 \begin{center}
  \begin{tabular}{|l|c|c|c|}
    \hline
    {\bf Search} & {\bf Period} & {\bf Lumi [$\pbi$]} &
      {\bf Evts. in DATA (MC)} \\
    \hline
    H1, di-muons & 1999-00 &  70.9 & --- \\
    \hline
    ZEUS, di-muons & 1997-00 & 105.2 & 200 ($213 \pm 11$)\\
    \hline
    H1, ``2e'' & 1994-97 & 115.2 & 105 ($118.2 \pm 12.8$) \\
    H1, ``3e''           &         &       &  16 ($ 21.6 \pm  3.0$) \\
    \hline
    ZEUS, ``2e'' & 1994-00 & 130.5 & 191 ($213  \pm 4$) \\
    ZEUS, ``3e''           &         &       & 26  ($34.7 \pm 0.5$)\\
    \hline
  \end{tabular}
 \end{center}
 \caption{Summary of the various analyses presented in this review.
   The data taking period, the integrated luminosity and the number of
   events (detected and expected) is shown.}
 \label{tab-summary}
\end{table}

\end{document}